\documentclass[12pt]{article}
\usepackage{amssymb,amsmath,epsfig}

\begin{document}

\title{\textbf{Fermions Tunneling from
Pleba$\acute{\textmd{n}}$ski-Demia$\acute{\textmd{n}}$ski
Black Holes}}

\author{M. Sharif \thanks{msharif.math@pu.edu.pk} and Wajiha Javed
\thanks{wajihajaved84@yahoo.com} \\
Department of Mathematics, University of the Punjab,\\
Quaid-e-Azam Campus, Lahore-54590, Pakistan.}

\date{}
\maketitle

\begin{abstract}
Hawking radiation spectrum via fermions tunneling is investigated
through horizon radii of
Pleba$\acute{\textmd{n}}$ski-Demia$\acute{\textmd{n}}$ski family of
black holes. To this end, we determine the tunneling probabilities
for outgoing and incoming charged fermion particles and obtain their
corresponding Hawking temperatures. The graphical behavior of
Hawking temperatures and horizon radii (cosmological and event
horizons) is also studied. We find consistent results with those
already available in literature.
\end{abstract}
{\bf Keywords:} Quantum tunneling; NUT solution.\\
{\bf PACS numbers:} 04.70.Dy; 04.70.Bw; 11.25.-w

\section{Introduction}

A visual representation of black hole (BH) illustrates that it
dissipates energy via radiation, hence compresses and finally
dissolves. Classically, BHs are stable objects, but due to emission
of quantum particles (which create quantum fluctuations) these
become unstable. Hawking \cite{2refHaw} suggested that BHs radiate
thermally and transmit energy/mass in the form of particles
radiation known as \emph{Hawking radiation}.

It has been interesting to explore quantum phenomenon of Hawking
radiation from BHs as a tunneling technique of emitting quantum
particles. Two different procedures are usually employed to compute
particles action by determining its imaginary component. Parikh and
Wilczek \cite{[5]} established the null geodesic approach by
following the work of Kraus and Wilczek \cite{krawil}, while the
second tunneling method is called Hamilton-Jacobi ansatz. Later,
Kerner and Mann \cite{rbmann1} extended the calculations of the
tunneling process for the spin-1/2 particles emission by using the
WKB approximation to the Dirac equation and calculated tunneling
probability for nonrotating BHs. Also, fermions tunneling is applied
to a general nonrotating BH and recovered the corresponding Hawking
temperature. The same authors \cite{rbmann} also investigated the
Hawking temperature through Kerr-Newman BH.

Dias and Lemos \cite{2refDias} analyzed pair of accelerated BHs in
de Sitter and anti-de Sitter backgrounds. Chen et al. \cite{WJ3}
investigated Hawking radiation spectrum via spin-1/2 particles
tunneling from rotating BHs in de Sitter space and recovered their
corresponding Hawking temperatures. Recently, the tunneling
probabilities from accelerating and rotating BHs have been
investigated for different particles \cite{ks1}. Also, the
thermodynamical properties of accelerating and rotating BHs with
Newman-Unti-Tamburino (NUT) parameter have been studied \cite{1-21}.

The effect of magnetic monopole (induced by NUT parameter)
hypothesis in general relativity was put forward by Dirac. He
suggested the innovative existence of magnetic monopole that was
neglected due to the failure to detect such object. Recently, the
new developments in relativistic quantum field theory has shed light
on it. Cot\u{a}escu and Visinescu \cite{rep1} investigated the Dirac
field in Taub-NUT background. Kerner and Mann \cite{rep5} obtained
the temperature of Taub-NUT-anti-de Sitter BHs by using
null-geodesic method and the Hamilton-Jacobi ansatz. Ali \cite{rep6}
investigated tunneling radiation characteristics from the hot
NUT-Kerr-Newman-Kasuya spacetime.

Li and Han \cite{rep2} extended the Kerner and Mann fermions
tunneling framework to study the tunneling of charged and magnetized
fermions from the RN BH with magnetic charges. Wang and Yang
\cite{rep3} studied Hawking radiation via charged fermions from the
NUT Kerr-Newman BH and recovered consistent Hawking temperature.
Xiao-Xiong and Qiang \cite{rep4} discussed tunneling of scalar and
Dirac particles from the Taub-NUT-AdS BH by using the
Hamilton-Jacobi method as well as Kerner and Mann tunneling
approach. The corresponding general form of the temperature of
scalar and Dirac particles is obtained.

We have explored few application of the tunneling phenomenon for
different BHs \cite{S1} by using the above mentioned methods. In a
recent paper \cite{S1a}, we have investigated some interesting
results for a group of BHs which exhibits a pair of charged NUT
accelerating and rotating BH solution. This paper extends the
tunneling phenomenon of charged fermions for the
Pleba$\acute{\textmd{n}}$ski-Demia$\acute{\textmd{n}}$ski (PD) class
of BHs which symbolizes a combination of charged NUT accelerating
and rotating BH solution with cosmological constant $\Lambda$.

The paper is planed as follows. Section \textbf{2} is devoted to
explain the basic equations for a PD class of BHs. In section
\textbf{3}, we provide Dirac equation in the framework of PD BHs and
evaluate the tunneling probabilities as well as the corresponding
temperatures across the horizon radii. Also, we evaluate a precise
construction of the particles action. Finally, we summarize the
results in the last section.

\section{Pleba$\acute{\textmd{n}}$ski-Demia$\acute{\textmd{n}}$ski
Family of Black Holes}

Black holes are extremely valuable objects conjectured by general
relativity \cite{[1]}. The research in this area has been broaden by
addition of different sources, e.g., electric and magnetic charges,
acceleration, rotation, cosmological constant as well as NUT
parameter in the usual mass of BH. Black hole solutions with these
extensions belong to type D class. This class of type D spacetimes
can be described by a metric proposed by
Pleba$\acute{\textmd{n}}$ski and Demia$\acute{\textmd{n}}$ski
\cite{[8]}. The PD metric can be reduced to the entire class of type
D BHs including a nonzero cosmological constant and electromagnetic
field by applying coordinate transformation in certain limits.

The PD metric can be interpreted by introducing two continuous
parameters that represent acceleration $\alpha$ and twist $\omega$
of the sources via rescaling. The twist is entirely expressed in
terms of angular velocity and NUT-like properties of the sources.
Using coordinate transformations in the modified form of the general
PD metric, $a$ (rotation parameter of Kerr-like BH) and $l$ (NUT
parameter) can be introduced, leading to the PD BHs \cite{NEW1}.
Some important BH subfamilies depend upon these parameters.

The class of BH solutions can be written in the form \cite{NEW1}
\begin{eqnarray}
\textmd{d}s^2&=&-\frac{1}{\Omega^2}\left\{\frac{Q}{\rho^2}
\left[\textmd{d}t-\left(a\sin^2\theta+4l\sin^2\frac{\theta}{2}
\right)\textmd{d}\phi\right]^2- \frac{\rho^2}{Q}
\textmd{d}r^2\right.\nonumber\\
&-&\left.\frac{\tilde{P}}{\rho^2}\left[a\textmd{d}t-\left(r^2+(a+l)^2\right)
\textmd{d}\phi\right]^2-\frac{\rho^2}{\tilde{P}}\sin^2\theta
\textmd{d}\theta^2\right\}\label{NN},
\end{eqnarray}
where
\begin{eqnarray}\nonumber
\Omega&=&1-\frac{\alpha}{\omega}(l+a\cos\theta)r,\quad
\rho^2=r^2+(l+a\cos\theta)^2,\nonumber\\
Q&=&(\omega^2k+e^2+g^2)-2Mr+\epsilon r^2-2\alpha\frac{n}{\omega}r^3-
\left(\alpha^2k+\frac{\Lambda}{3}\right)r^4,\nonumber\\
\tilde{P}&=&\sin^2\theta(1-a_3\cos\theta-a_4\cos^2\theta)=P\sin^2\theta,\nonumber\\
a_3&=&2\alpha\frac{a}{\omega}M-4\alpha^2\frac{
al}{\omega^2}(\omega^2k+e^2+g^2)-4\frac{\Lambda}{3}al,\nonumber\\
a_4&=&-\alpha^2\frac{a^2}{\omega^2}(\omega^2k+e^2+g^2)-\frac{\Lambda}{3}a^2,\nonumber\\
\epsilon&=&\frac{\omega^2k}{a^2-l^2}+4\alpha\frac{
l}{\omega}M-(a^2+3l^2)\left[\frac{\alpha^2}{\omega^2}(\omega^2k+e^2+g^2)+
\frac{\Lambda}{3}\right],\nonumber\\
n&=&\frac{\omega^2kl}{a^2-l^2}-\alpha\frac{a^2-l^2}{\omega}M+
l(a^2-l^2)\left[\frac{\alpha^2}{\omega^2}(\omega^2k+e^2+g^2)+
\frac{\Lambda}{3}\right],\nonumber\\
k&=&\left(\frac{\omega^2}{a^2-l^2}+3\alpha^2l^2\right)^{-1}\left[1+2\alpha\frac{
l}{\omega}M-3\alpha^2\frac{l^2}{\omega^2}(e^2+g^2)-l^2\Lambda\right].\nonumber
\end{eqnarray}
Here, the arbitrary parameters $M,~e,~g,~\Lambda,~a,~l$ and $\alpha$
vary independently, while parameter $\omega$ varies dependently (in
some sub-cases), and $\epsilon,~n,~k$ are arbitrary real parameters.
All parameters in PD BHs except $\Lambda,~e,~g$ do not have their
physical interpretation, but have their usual physical significance
in certain sub-cases. Electric and magnetic charges of the source
are denoted by $e$ and $g$, respectively, while $M$ is the source
mass and $n$ is the PD parameter. Notice that this class of BH
involves acceleration $\alpha$ and twisting behavior $\omega$.

Generally, the NUT parameter is analogous with the gravitomagnetic
monopole parameter of the central mass, or a twisting property of
the surrounding spacetime but its exact physical meaning could not
be found. If $l>a$ (for this BH), the spacetime will be free from
curvature singularities and the resulting solution is characterized
by the NUT-like solution. However, if the rotation parameter governs
the NUT parameter, i.e., $a>l$, the solution corresponds to the
Kerr-like and forms a ring curvature singularity. Such singularity
structure does not depend on cosmological constant. The cosmological
constant has dynamical nature which provides expanding solutions
when $\Lambda>0$ (de Sitter space) and provides asymptotic regions
with constant curvature when $\Lambda<0$ (anti-de Sitter space).
Here, the PD BH solutions belong to the de Sitter family of
solutions and PD metric reduces to the expanding BHs with
$\Lambda>0$.

Kerr-Newman solution with NUT parameter in de Sitter space is
obtained for $\alpha=0$ and $\omega^2k=(1-l^2\Lambda)(a^2-l^2)$. In
this case, $\omega$ is related to both $a$ and $l$. Thus,
$M,~e,~g,~l,a$ vary independently, while $\omega$ depends on nonzero
value of rotation parameters $l$ or $a$. It can be re-expressed by
choosing $a$ and $l$. For $l=0$, this leads to the Kerr-Newman
accelerating de Sitter pair of BHs, while $\alpha=0$ leads to the
Kerr-Newman BH in de Sitter space and $a=0$ yields the RN BH. In
addition, if $e=0=g$, we have Schwarzschild BH. Thus, the metric
(\ref{NN}) for the generalized BHs represents complete family of
BHs. For $a=0$, this leads to the C-metric having charge and
cosmological constant, consequently for $\Lambda=0$ we retrieve the
exact charged shape of the C-metric.

The metric (\ref{NN}) can be expressed in another more suitable form
\begin{equation}
\textmd{d}s^2=-f(r,\theta)\textmd{d}t^2
+\frac{\textmd{d}r^2}{g(r,\theta)}+\Sigma(r,\theta)
\textmd{d}\theta^2+K(r,\theta)\textmd{d}
\phi^2-2H(r,\theta)\textmd{d}t\textmd{d}\phi,
\end{equation}
where $f(r,\theta),~g(r,\theta),~\Sigma(r,\theta),~K(r,\theta)$ and
$H(r,\theta)$ can be defined as follows
\begin{eqnarray}\nonumber
f(r,\theta)&=&\left(\frac{Q-Pa^2\sin^2\theta}{\rho^2\Omega^2}
\right),\quad g(r,\theta)=\frac{Q\Omega^2}
{\rho^2},\quad\Sigma(r,\theta)=\frac{\rho^2}{\Omega^2P},\\\nonumber
K(r,\theta)&=&\frac{1}{\rho^2\Omega^2}\left[\sin^2\theta
P[r^2+(a+l)^2]^2-Q(a\sin^2\theta+4l\sin^2
\frac{\theta}{2})^2\right],\\
H(r,\theta)&=&\frac{1}{\rho^2\Omega^2}\left[\sin^2\theta
Pa[r^2+(a+l)^2]-Q(a\sin^2
\theta+4l\sin^2\frac{\theta}{2})\right].\nonumber
\end{eqnarray}
The four-vector potential for these BHs can be determined as
\cite{acc1}
\begin{eqnarray}\nonumber
A_\mu&=&\frac{1}{a[r^2+(l+a\cos\theta)^2]}
[-er[a\textmd{d}t-\textmd{d}\phi\{(l+a)^2
-(l^2+a^2\cos^2\theta\nonumber\\
&+&2la\cos\theta)\}]
-g(l+a\cos\theta)[a\textmd{d}t-\textmd{d}\phi\{r^2+(l+a)^2\}]].\nonumber
\end{eqnarray}
The horizons are found for
$g(r,\theta)=\frac{\Delta(r)}{\Sigma(r,\theta)}=0$ \cite{rbmann},
where $\Delta(r)=\frac{Q}{P}$. This implies that $\Delta(r)=0=Q$,
yielding the horizon radii
\begin{eqnarray}
r_{1\pm}=-A'-B'\pm0.5\sqrt{D-E},\label{New4}\\
r_{2\pm}=-A'+B'\pm0.5\sqrt{D+E}.\label{New5}
\end{eqnarray}

In the above equations, BH horizons are denoted by $r_+$ (outer) and
$r_-$ (inner). The values $A',~B',~D$ and $E$ are defined as
\begin{eqnarray}
A'&=&0.25\frac{B}{C},\nonumber\\
B'&=&0.5\left[\left\{0.25\frac{B^2}{C^2}+0.666667\frac{\epsilon}{C}+5.03968(-AC+0.0833333\epsilon^2
\right.\right.\nonumber\\
&-&0.5BM)\left[C\left\{-27AB^2-72AC\epsilon-2\epsilon^3+18B\epsilon
M+108CM^2\right.\right.\nonumber\\
&+&\left[-4(-12AC+\epsilon^2-6BM)^3+(-27AB^2-72AC\epsilon-
2\epsilon^3\right.\nonumber\\&+&18B\epsilon
M\left.\left.\left.+108CM^2)^2\right]^\frac{1}{2}\right\}^\frac{1}{3}\right]^{-1}+
\frac{0.264567}{C}\left\{-27AB^2-72AC\epsilon\right.\nonumber\\&-&2\epsilon^3+18B\epsilon
M+108CM^2+\left[-4(-12AC+\epsilon^2-6BM)^3\right.\nonumber\\&+&
\left.\left.\left.\left.(-27AB^2-72AC\epsilon-2\epsilon^3+18B\epsilon
M+108CM^2)^2\right]^\frac{1}{2}\right\}^\frac{1}{3}\right\}\right]^{\frac{1}{2}},\nonumber
\end{eqnarray}
\begin{eqnarray}
D&=&\left[0.5\frac{B^2}{C^2}+1.33333\frac{\epsilon}{C}-5.03968(-AC+0.0833333\epsilon^2
\right.\nonumber\\
&-&0.5BM)\left[C\left\{-27AB^2-72AC\epsilon-2\epsilon^3+18B\epsilon
M+108CM^2\right.\right.\nonumber\\
&+&\left[-4(-12AC+\epsilon^2-6BM)^3+(-27AB^2-72AC\epsilon-
2\epsilon^3\right.\nonumber\\
&+&18B\epsilon
M\left.\left.\left.+108CM^2)^2\right]^\frac{1}{2}\right\}^\frac{1}{3}\right]^{-1}-
\frac{0.264567}{C}\left\{-27AB^2-72AC\epsilon\right.\nonumber\\
&-&2\epsilon^3+18B\epsilon
M+108CM^2+\left[-4(-12AC+\epsilon^2-6BM)^3\right.\nonumber\\
&+&\left.\left.\left.(-27AB^2-72AC\epsilon-2\epsilon^3+18B\epsilon
M+108CM^2)^2\right]^\frac{1}{2}\right\}^\frac{1}{3}\right],\nonumber\\
E&=&\frac{0.125}{\acute{B}}\left(-\frac{B^3}{C^3}-
4\frac{B\epsilon}{C^2}-16\frac{M}{C}\right),\nonumber
\end{eqnarray}
while $A,~B$ and $C$ can be written as
\begin{equation*}
A=(\omega^2k+e^2+g^2),\quad B=\frac{2\alpha n}{\omega},\quad
C=\alpha^2k+\frac{\Lambda}{3},
\end{equation*}
satisfying the following condition
\begin{eqnarray}
&&\left[-4(-12AC+\epsilon^2-6BM)^3+
(-27AB^2\right.\nonumber\\
&-&72\left.AC\epsilon-2\epsilon^3+18B\epsilon
M+108CM^2)^2\right]^\frac{1}{2}>0.\nonumber
\end{eqnarray}

The expression of angular velocity at the BH horizons can be defined
as
\begin{equation*}
\Omega_H=\frac{H(r_+,\theta)}{K(r_+,\theta)}=\frac{a}{r_+^2+(a+l)^2},
\end{equation*}
where $r_+$ corresponds to $r_{1+}$ and $r_{2+}$. The inverse
function of $f(r,\theta)$ is
\begin{equation*}
F(r,\theta)=f(r,\theta)+\frac{H^2(r,\theta)}{K(r,\theta)}.
\end{equation*}
For these BHs, the above expression becomes
\begin{equation*}
F(r,\theta)=\frac{PQ\sin^2\theta\rho^2}{\Omega^2[\sin^2\theta
P[r^2+(a+l)^2]^2-Q(a\sin^2\theta+4l\sin^2\frac{\theta}{2})^2]}.
\end{equation*}
In terms of $\Delta(r)$ and $\Sigma$, we can write the inverse
function of $f(r,\theta)$ as
\begin{equation*}
F(r,\theta)=\frac{P^2\Delta(r)\Sigma(r,\theta)}{[r^2+(a+l)^2]^2
-\Delta(r)\sin^2\theta[a+\frac{2l}{1+\cos\theta}]^2}.
\end{equation*}

\section{Charged Particles Tunneling}

In order to study charged fermions tunneling of mass $m$ from a
class of PD BHs, we consider the Dirac equation in covariant form as
\cite{rn}
\begin{equation}
\iota\gamma^\mu\left(D_\mu-\frac{\iota
q}{\hbar}A_\mu\right)\Psi+\frac{m}{\hbar}\Psi=0,\quad
\mu=0,1,2,3\label{2'}
\end{equation}
where $q$ is electric charge, $A_\mu$ is the four-potential, $\Psi$
is the wave function and
\begin{equation*}
D_\mu=\partial_\mu+\Omega_\mu,\quad
\Omega_\mu=\frac{1}{2}\iota\Gamma^{\alpha\beta}_
\mu\Sigma_{\alpha\beta},\quad\Sigma_{\alpha\beta}=
\frac{1}{4}\iota[\gamma^\alpha,\gamma^\beta].
\end{equation*}
Dirac matrices \cite{ks1} imply that
$[\gamma^\alpha,\gamma^\beta]=0$ for $\alpha=\beta$ and
$[\gamma^\alpha,\gamma^\beta]=-[\gamma^\beta,\gamma^\alpha]$ for
$\alpha\neq\beta$. Consequently, Eq.(\ref{2'}) reduces to
\begin{equation}
\iota\gamma^\mu\left(\partial_\mu-\frac{\iota
q}{\hbar}A_\mu\right)\Psi+\frac{m}{\hbar}\Psi=0.\label{2}
\end{equation}

The spinor wave function $\Psi$ (related to the particle's action)
has two spin states: in $+$ve $r$-direction (spin-up) and in $-$ve
$r$-direction (spin-down). For the spin-up and spin-down particle's
solution, we assume \cite{rbmann1}
\begin{eqnarray}\nonumber
\Psi_\uparrow(t,r,\theta,\phi)&=&\left[\begin{array}{c}
A(t,r,\theta,\phi)\\0\\
B(t,r,\theta,\phi)\\0
\end{array}\right]\exp\left[\frac{\iota}{\hbar}
I_\uparrow(t,r,\theta,\phi)
\right],\label{1}\\
\Psi_\downarrow(t,r,\theta,\phi)&=&\left[\begin{array}{c}
0\\C(t,r,\theta,\phi)\\0\\
D(t,r,\theta,\phi)
\end{array}\right]\exp\left[\frac{\iota}{\hbar}
I_\downarrow(t,r,\theta,\phi) \right],\nonumber
\end{eqnarray}
where $I_{\uparrow/\downarrow}$ denote the emitted spin-up/spin-down
particle's action, respectively. Here, we deal with only spin-up
particles, while calculations for spin-down particles is similar as
above.

The particle's action through Hamilton-Jacobi ansatz \cite{krawil,
rbmann1} is
\begin{equation}
I_\uparrow=-Et+J\phi+W(r,\theta),\label{WWWW}
\end{equation}
where $E,~J,~W$ are the energy, angular momentum and arbitrary
function, respectively. Using this ansatz in the Dirac equation with
$\iota A=B,~\iota B=A$ and Taylor's expansion of $F(r,\theta)$ near
the horizon $r_+$, it follows that
\begin{eqnarray}
&-&B\left[\frac{-E+\Omega_HJ+\frac{qer}{[r_+^2+(a+l)^2]}}
{\sqrt{(r-r_+)\partial_rF(r_+,\theta)}}
+\sqrt{(r-r_+)\partial_rg(r_+,\theta)}(\partial_rW)\right]+mA=0,\nonumber\\\label{3}\\
&-&B\left[\sqrt{\frac{P\Omega^2
(r_+,\theta)}{\rho^2(r_+,\theta)}} (\partial_\theta W)\nonumber\right.\\
&+&\left.\frac{\iota\rho(r_+,\theta)
\Omega(r_+,\theta)}{\sqrt{\sin^2\theta
P[r^2+(a+l)^2]^2-Q(a\sin^2\theta+4l\sin^2\frac{\theta}{2})^2}}\right.\nonumber\\
&\times&\left.\left\{J-q
\left[\frac{er[(l+a)^2-(l^2+a^2\cos^2\theta+2la\cos\theta)]}
{a[r^2+(l+a\cos\theta)^2]}\right.
\right.\right.\nonumber\\
&+&\left.\left.\left.\frac{g(l+a\cos\theta)[r^2+(l+a)^2]}{a[r^2+(l+a\cos\theta)^2]}
\right]\right\}\right]=0,\label{4}
\end{eqnarray}
\begin{eqnarray}
&&A\left[\frac{-E+\Omega_HJ+\frac{qer}{[r^2+(a+l)^2]}}
{\sqrt{(r-r_+)\partial_rF(r_+,\theta)}}-\sqrt{(r-r_+)
\partial_rg(r_+,\theta)}(\partial_rW)\right]+mB=0,\nonumber\\\label{5}\\
&-&A\left[\sqrt{\frac{P\Omega^2(r_+,\theta)}{\rho^2(r_+,\theta)}}
(\partial_\theta W)\nonumber\right.\\
&+&\left.\frac{\iota\rho(r_+,\theta)
\Omega(r_+,\theta)}{\sqrt{\sin^2\theta
P[r^2+(a+l)^2]^2-Q(a\sin^2\theta+4l\sin^2\frac{\theta}{2})^2
}}\right.\nonumber\\&\times&\left.\left\{J-q \left[\frac{er[(l+a)^2-
(l^2+a^2\cos^2\theta+2la\cos\theta)]}{a[r^2+(l+a\cos\theta)^2]}
\right. \right.\right.\nonumber\\&+&\left.\left.\left.\frac{
g(l+a\cos\theta)[r^2+(l+a)^2]}{a[r^2+(l+a\cos\theta)^2]}\right]
\right\}\right]=0.\label{6}
\end{eqnarray}
The arbitrary function $W(r,\theta)$ can be separated as follows
\cite{rbmann}
\begin{equation}
W(r,\theta)=R(r)+\Theta(\theta).\label{7}
\end{equation}

Firstly, we deal with Eqs.(\ref{3})-(\ref{6}) for massless ($m=0$)
case. Consequently, Eqs.(\ref{3}) and (\ref{5}) reduce to
\begin{eqnarray}
\frac{-E+\Omega_HJ+\frac{qer}{[r_+^2+(a+l)^2]}}
{\sqrt{(r-r_+)\partial_rF(r_+,\theta)}}+
\sqrt{(r-r_+)\partial_rg(r_+,\theta)}R^\prime(r)=0,\\
\frac{-E+\Omega_HJ+\frac{qer}{[r^2+(a+l)^2]}}
{\sqrt{(r-r_+)\partial_rF(r_+,\theta)}}-\sqrt{(r-r_+)
\partial_rg(r_+,\theta)}R^\prime(r)=0.
\end{eqnarray}
For $r_+=r_{1+}$, the above equations imply that
\begin{eqnarray}
R^\prime(r)=R^\prime_+(r)=-R^\prime_-(r)&=&
\left[\frac{[r_{1+}^2+(a+l)^2]}{\left(r-r_{1+}\right)\left(r_{1+}-r_{2+}\right)
\left(r_{1+}-r_{2-}\right)}\right.\nonumber\\
&\times&\left.\frac{\left(E-\Omega_HJ-\frac{qer_{1+}}
{[r_{1+}^2+(a+l)^2]}\right)}{\left(r_{1+}-r_{1-}\right)}\right],\label{AW}
\end{eqnarray}
where $R_+$ and $R_-$ correspond to the outgoing and incoming
solutions, respectively. This equation represents the pole at the
horizon, $r=r_{1+}$.

Integrating Eq.(\ref{AW}) around the pole \cite{S1a}, we obtain
\begin{eqnarray}
R_+(r)=-R_-(r)=\left[\frac{\pi
\iota\left[r_{1+}^2+(a+l)^2\right]\left(E-\Omega_HJ-\frac{qer_{1+}}
{[r_{1+}^2+(a+l)^2]}\right)}{\left(r_{1+}-r_{2+}\right)
\left(r_{1+}-r_{2-}\right)\left(r_{1+}-r_{1-}\right)}\right].
\end{eqnarray}
The imaginary parts of $R_+$ and $R_-$ yield
\begin{eqnarray}
\textmd{Im}R_+=-\textmd{Im}R_-=\left[\frac{\pi\left[r_{1+}^2+(a+l)^2\right]
\left(E-\Omega_HJ-\frac{qer_{1+}}
{\left[r_{1+}^2+(a+l)^2\right]}\right)}{\left(r_{1+}-r_{2+}\right)
\left(r_{1+}-r_{2-}\right)\left(r_{1+}-r_{1-}\right)}\right].
\end{eqnarray}
Thus, the outgoing particle's tunneling probability is
\begin{eqnarray}
\Gamma&=&\frac{\textmd{Prob}[\textmd{out}]}
{\textmd{Prob}[\textmd{in}]}=
\frac{\exp[-2(\textmd{Im}R_++\textmd{Im}\Theta)]}
{\exp[-2(\textmd{Im}R_-+\textmd{Im}\Theta)]}=
\exp[-4\textmd{Im}R_+]\nonumber\\
&=&\exp\left[\frac{-4\pi\left[r_{1+}^2+(a+l)^2\right]\left(E-\Omega_HJ-\frac{qer_{1+}}
{\left[r_{1+}^2+(a+l)^2\right]}\right)}{\left(r_{1+}-r_{2+}\right)
\left(r_{1+}-r_{2-}\right)\left(r_{1+}-r_{1-}\right)}\right].\label{12}
\end{eqnarray}
Using the WKB approximation, $\Gamma$ is given in terms of classical
action $I$ of charged particles up to leading order in $\hbar$.
Thus, for calculating the Hawking temperature, we expand the action
in terms of particles energy $E$, i.e., $2I=\beta E+O(E^2)$ so that
the Hawking temperature is recovered at linear order
\begin{equation}
\Gamma\sim\exp[-2I]\simeq\exp[-\beta E].
\end{equation}
This shows that the emission rate in the tunneling approach, up to
first order in $E$, retrieves the Boltzmann factor of the form
$\exp[-\beta E]$ with $\beta=\frac{1}{T_H}$ \cite{s2}. The
higher-order terms represent self-interaction effects resulting from
the energy conservation.

The required Hawking temperature at horizon $r_{1+}$ can be written
as
\begin{equation}
T_H=\left[\frac{\left(r_{1+}-r_{1-}\right)\left(r_{1+}-r_{2+}\right)\left(r_{1+}-r_{2-}\right)}
{4\pi[r_{1+}^2+(a+l)^2]}\right].\label{New1}
\end{equation}
\begin{figure}\center
\epsfig{file=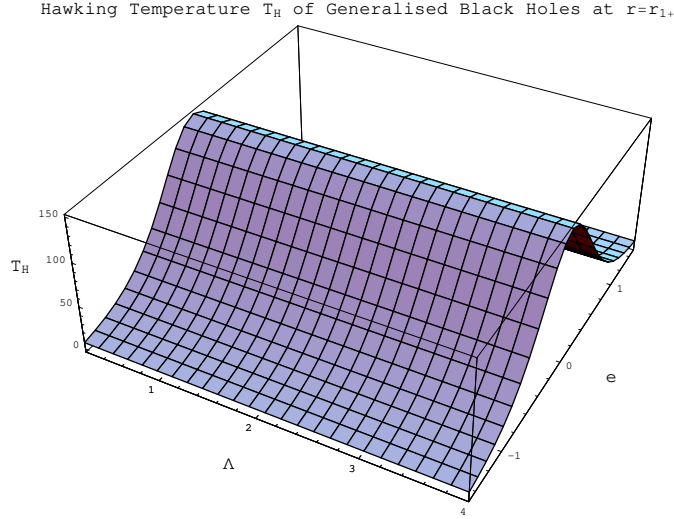, width=0.65\linewidth}\\
\caption{Hawking temperature $T_H$ at $r_{1+}$ versus cosmological
constant $\Lambda$ and electric charge $e$}
\end{figure}
When $l=0,~k=1$ and $\Lambda=0$ in Eq.(\ref{New1}), the Hawking
temperature of the accelerating and rotating BHs, electric and
magnetic charges is recovered \cite{ks1}. For $\alpha=0$, it reduces
to the temperature of non-accelerating BHs \cite{1-21}, while
$l=0,~k=1,~\alpha=0$, gives Hawking temperature of the Kerr-Newman
BH \cite{rbmann}, which further reduces to the temperature of the RN
BH (for $a=0$). Finally, in the absence of charge, it exactly
becomes the Hawking temperature of the Schwarzschild BH \cite{A}. In
case of massive particles ($m\neq0$), following the same steps, we
can obtain the same temperature. Thus the behavior will be same for
both massive and massless particles near the BH horizon. For
$\omega=1.25, M=10, \alpha=10, g=1, a=22, l=100$ (based on the
cosmological constant $\Lambda$ and electric charge $e$), the
graphical representation of Hawking temperature (\ref{New1}) (at
$r=r_{1+}$) and the corresponding horizon $r=r_{1+}$ (\ref{New4}) of
PD BHs is shown in Figures \textbf{1} and \textbf{2}, respectively.
\begin{figure}\center
\epsfig{file=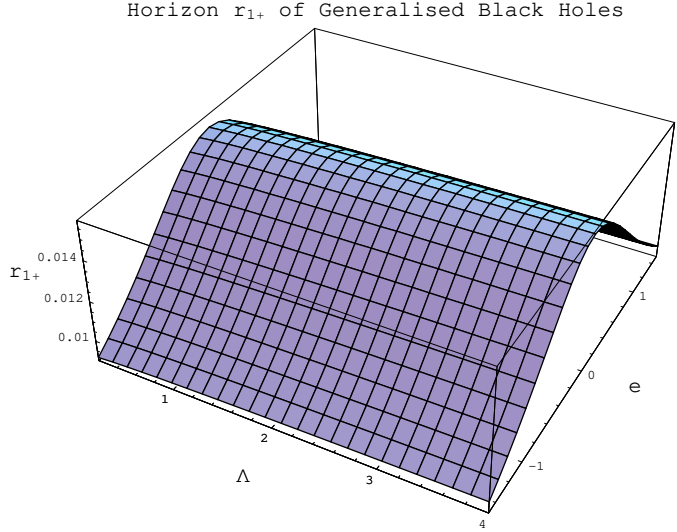, width=0.65\linewidth}\\
\caption{Horizon radius $r_{1+}$ versus cosmological constant
$\Lambda$ and electric charge $e$}
\end{figure}

Now, we explore the tunneling probability of charged massive and
massless fermions from the horizon $r_{2+}$ given in Eq.(\ref{New5})
by using the similar process. The corresponding set of
Eqs.(\ref{3})-(\ref{6}) for the outgoing and incoming fermions,
respectively, yield
\begin{eqnarray}
R_+(r)=-R_-(r)=\left[\frac{\pi\iota[r_{2+}^2+(a+l)^2]\left(E-\Omega_\alpha
J-\frac{qer_{2+}}{[r_{2+}^2+(a+l)^2]}\right)}{\left(r_{2+}-r_{1+}\right)
\left(r_{2+}-r_{1-}\right)\left(r_{2+}-r_{2-}\right)}\right].
\end{eqnarray}
The probability for particles which tunnel through horizon will be
\begin{eqnarray}
\Gamma&=&\exp\left[\frac{-4\pi[r_{2+}^2+(a+l)^2]\left(E-\Omega_H
J-\frac{qer_{2+}}{[r_{2+}^2+(a+l)^2]}\right)}
{\left(r_{2+}-r_{2-}\right)\left(r_{2+}-r_{1+}\right)
\left(r_{2+}-r_{1-}\right)}\right].\label{12A}
\end{eqnarray}
Consequently, the corresponding temperature value (at $r_{2+}$) is
\begin{equation}
T_H=\left[\frac{\left(r_{2+}-r_{2-}\right)\left(r_{2+}-r_{1+}\right)
\left(r_{2+}-r_{1-}\right)}
{4\pi[r_{2+}^2+(a+l)^2]}\right].\label{NewWWW}
\end{equation}
For $\omega=9,~M=0.9,~\alpha=0.1,~g=15,~a=100,~l=10$, Figures
\textbf{3} and \textbf{4} show that the Hawking temperature
(\ref{NewWWW}) at horizon radius $r_{2+}$ (\ref{New5}) always
remains positive for the above mentioned parameters. The horizon
radius $r_{2+}$ and Hawking temperature $T_H$ vanish as $\Lambda$
decreases and approach to zero for all $e$.
\begin{figure}\center
\epsfig{file=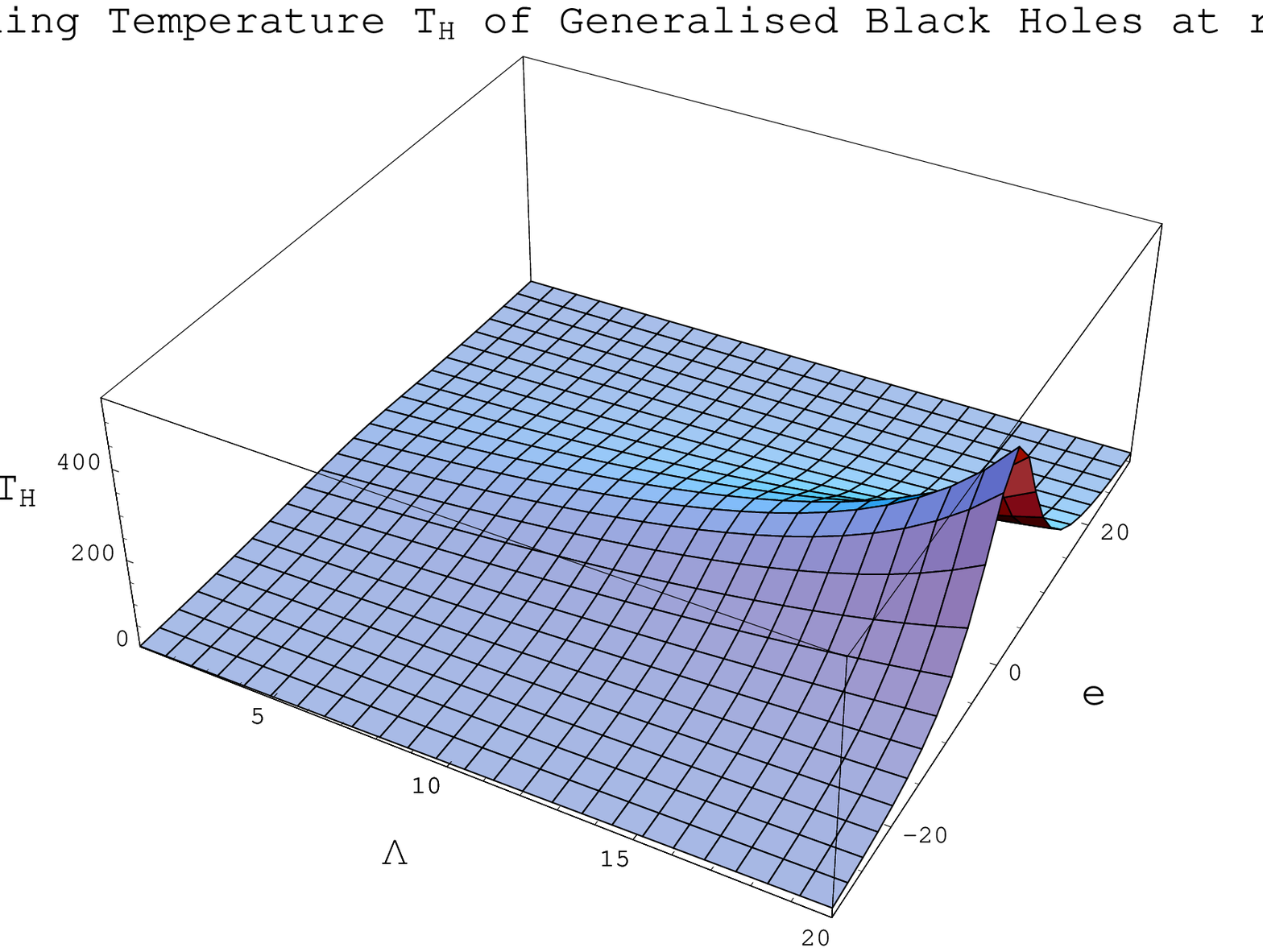, width=0.65\linewidth}\\
\caption{Hawking temperature $T_H$ at $r_{2+}$ versus cosmological
constant $\Lambda$ and electric charge $e$}
\epsfig{file=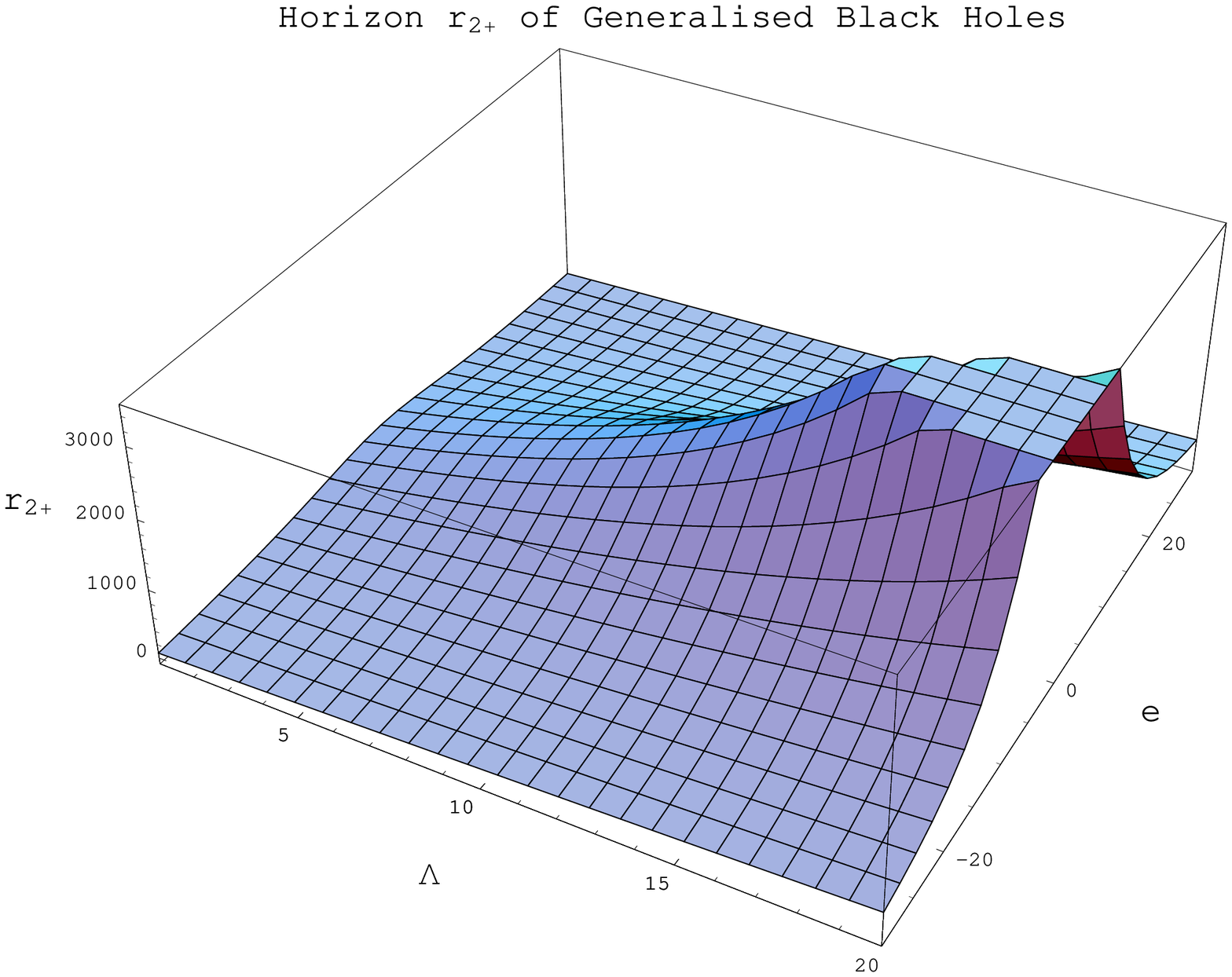, width=0.65\linewidth}\\
\caption{Horizon radius $r_{2+}$ versus cosmological constant
$\Lambda$ and electric charge $e$}
\end{figure}

In general, both horizon temperatures will differ, but there is a
set of parameters for which both temperatures have similar behavior
at $r_{1+}$ and $r_{2+}$ given in Figures \textbf{5} and \textbf{6},
respectively. The required set of parameters is given by
$\omega=25,~M=50,~\alpha=12.5,~g=.01,~a=20,~l=100$.
\begin{figure}\center
\epsfig{file=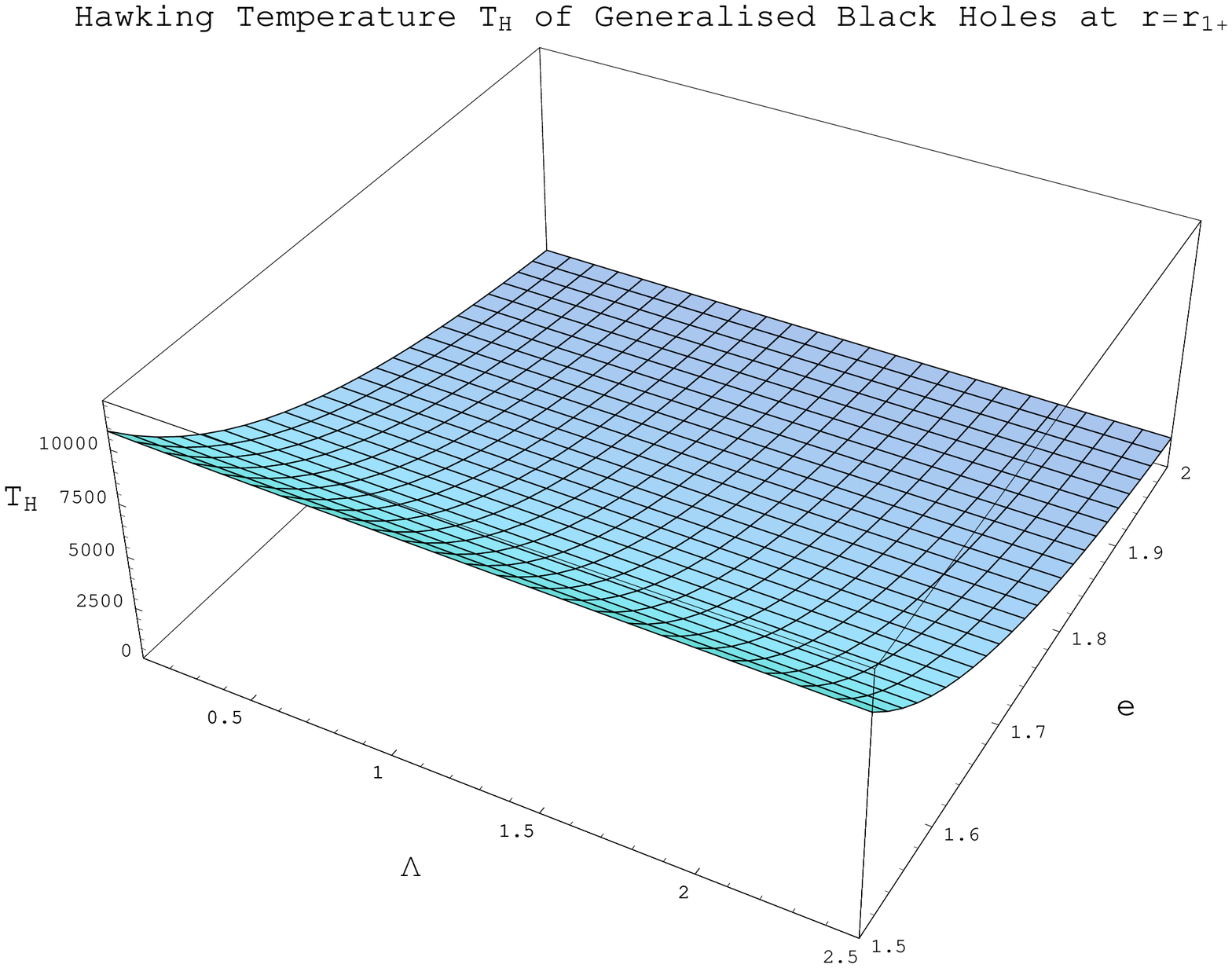, width=0.65\linewidth}\\
\caption{Hawking temperature $T_H$ at $r_{1+}$ versus cosmological
constant $\Lambda$ and electric charge $e$}
\epsfig{file=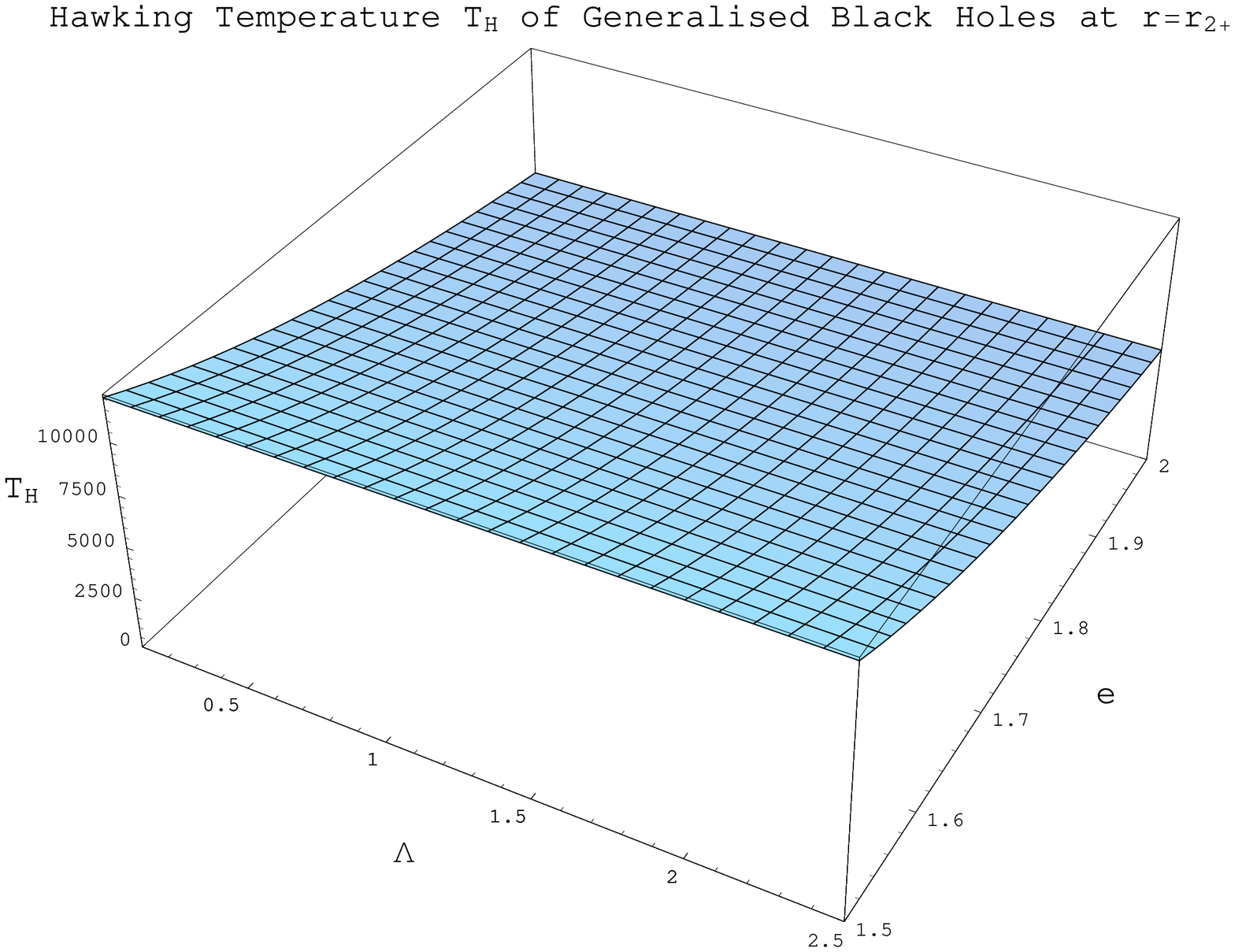, width=0.65\linewidth}\\
\caption{Hawking temperature $T_H$ at $r_{2+}$ versus cosmological
constant $\Lambda$ and electric charge $e$}
\end{figure}

\subsection{Action for the Emitted Particles}

We evaluate particle's action $I_\uparrow$ by using
Eqs.(\ref{3})-(\ref{6}). For outgoing particles, Eqs.(\ref{3}) and
(\ref{7}) can be expressed as follows
\begin{equation}
R^\prime(r)=\frac{mA}{B\sqrt{(r-r_+)
\partial_rg(r_+,\theta)}}-\frac{-E+\Omega_HJ+
\frac{qer_+}{[r_+^2+(a+l)^2]}}{(r-r_+)
\sqrt{\partial_rF(r_+,\theta)\partial_rg(r_+,\theta)}}.
\end{equation}
Integration with respect to $r$ provides
\begin{eqnarray}
R(r)=R_+(r)=\int\frac{mA}{B\sqrt{(r-r_+)
\partial_rg(r_+,\theta)}}\textmd{d}r\nonumber\\
-\frac{\left(-E+\Omega_HJ+ \frac{qer_+}{[r_+^2+(a+l)^2]}\right)}{
\sqrt{\partial_rF(r_+,\theta)\partial_rg(r_+,\theta)}}\ln(r-r_+).\label{WJ}
\end{eqnarray}
Similarly, in case of incoming particles, Eq.(\ref{5}) yields
\begin{eqnarray}
R(r)=R_-(r)=\int\frac{mB}{A\sqrt{(r-r_+)
\partial_rg(r_+,\theta)}}\textmd{d}r\nonumber\\+
\frac{\left(-E+\Omega_HJ+\frac{qer_+}{[r_+^2+(a+l)^2]}\right)}{
\sqrt{\partial_rF(r_+,\theta)\partial_rg(r_+,\theta)}}\ln(r-r_+).
\end{eqnarray}
Using Eq.(\ref{7}), we can write from Eqs.(\ref{4}) or (\ref{6}) as
\begin{eqnarray}
&&\sqrt{\frac{P\Omega^2(r_+,\theta)}{\rho^2(r_+,\theta)}}
\partial_\theta\Theta+\frac{\iota\rho(r_+,\theta)
\Omega(r_+,\theta)}{\sqrt{\sin^2\theta
P[r_+^2+(a+l)^2]^2}}\nonumber\\&\times&\left[J
-q\left\{\frac{er_+[(l+a)^2-
(l^2+a^2\cos^2\theta+2la\cos\theta)]}{a[r_+^2+(l+a\cos\theta)^2]}
\nonumber\right.\right.\\&+&\left.\left.\frac{
g(l+a\cos\theta)[r_+^2+(l+a)^2]}{a[r_+^2+(l+a\cos\theta)^2]}
\right\}\right]=0.
\end{eqnarray}

Inserting $\rho$ and $P$, after some manipulation, it follows that
\begin{eqnarray}
\partial_\theta\Theta&=&\frac{\iota a\sin\theta[aJ+qer_+]}{[r_+^2+(a+l)^2]}
\left[1-2\cos\theta\left\{\alpha
\frac{a}{\omega}M\right.\right.\nonumber\\&-&\left.2\alpha^2l
\frac{a}{\omega^2}(\omega^2k+e^2+g^2)-\frac{2\Lambda}{3}
al\right\}\nonumber\\&-&\left.\cos^2\theta\left\{-\alpha^2
\frac{a^2}{\omega^2}(\omega^2k+e^2+g^2)-\frac{\Lambda}{3}a^2\right\}\right]^{-1}
\nonumber\\&+&\frac{-\iota aJ+\iota
qg(l+a\cos\theta)}{a\sin\theta}\left[1-2\cos\theta\left\{\alpha
\frac{a}{\omega}M\right.\right.\nonumber\\&-&\left.2\alpha^2l
\frac{a}{\omega^2}(\omega^2k+e^2+g^2)-\frac{2\Lambda}{3}al\right\}\nonumber\\
&-&\left.
\cos^2\theta\left\{-\alpha^2\frac{a^2}{\omega^2}(\omega^2k+e^2+g^2)-
\frac{\Lambda}{3}a^2\right\}\right]^{-1} \nonumber\\&+&\frac{\iota
2l
(1-\cos\theta)[aJ+qer_+]}{\sin\theta[r_+^2+(a+l)^2]}\left[1-2\cos\theta
\left\{\alpha\frac{a}{\omega}M\right.\right.\nonumber\\
&-&\left.2\alpha^2l
\frac{a}{\omega^2}(\omega^2k+e^2+g^2)-\frac{2\Lambda}{3}al\right\}
\nonumber\\&-&\left.\cos^2\theta\left\{-\alpha^2
\frac{a^2}{\omega^2}(\omega^2k+e^2+g^2)-\frac{\Lambda}{3}a^2\right\}\right]^{-1}.\nonumber
\end{eqnarray}
Integrating with respect to $\theta$, we have
\begin{equation}
\Theta=\frac{\iota a[aJ+qer_+]}{[r_+^2+(a+l)^2]}I_1+I_2+
\frac{2l\iota[aJ+qer_+]}{[r_+^2+(a+l)^2]}I_3,\label{10}
\end{equation}
where $I_1,~I_2$ and $I_3$ are given as follows
\begin{eqnarray*}
I_1&=&\int\left[\frac{\sin\theta}{1-2\tilde{A}\cos\theta-\tilde{B}\cos^2\theta}\right]
\textmd{d}\theta,\\
I_2&=&\int\left[\frac{\iota qg(l+a\cos\theta)-\iota a
J}{a\sin\theta[1-2\tilde{A}\cos\theta-\tilde{B}\cos^2\theta]}\right]\textmd{d}\theta,
\nonumber\\
I_3&=&\int\left[\frac{1-\cos\theta}{\sin\theta[1-2\tilde{A}\cos\theta-\tilde{B}\cos^2\theta]}
\right]\textmd{d}\theta,
\end{eqnarray*}
and
\begin{eqnarray}
\tilde{A}&=&\left[\alpha\frac{a}{\omega}M-2\alpha^2l\frac{a}{\omega^2}(\omega^2k+e^2+g^2)-
\frac{2\Lambda}{3}al\right],\nonumber\\
\tilde{B}&=&\left[-\alpha^2\frac{a^2}{\omega^2}(\omega^2k+e^2+g^2)-\frac{\Lambda}{3}a^2\right].\nonumber
\end{eqnarray}

Solving these integrals, we obtain after some algebra
\begin{eqnarray}
I_1&=&\left[\frac{1}{2\sqrt{\tilde{A}^2+\tilde{B}}}\ln\left[\frac{1-x(\tilde{A}+
\sqrt{\tilde{A}^2+\tilde{B}})}{1-x(\tilde{A}-\sqrt{\tilde{A}^2+\tilde{B}})}\right]\right],\\
I_2&=&L_1\ln\left[\frac{1-x(\tilde{A}+
\sqrt{\tilde{A}^2+\tilde{B}})}{1-x(\tilde{A}-\sqrt{\tilde{A}^2+\tilde{B}})}\right]+L_2
\ln\left[1-2\tilde{A}x-\tilde{B}x^2\right]\nonumber\\&+&L_3
\ln[1-\cos\theta]+L_4\ln[1+\cos\theta],
\end{eqnarray}
where
\begin{eqnarray*}
L_1&=&\left[\frac{1}{2\sqrt{\tilde{A}^2+\tilde{B}}[(1-\tilde{B})^2-4\tilde{A}^2]}\right]\left[\iota
J(2\tilde{A}^2-\tilde{B}^2+\tilde{B})\right.\nonumber\\
&+&\left.\iota
qg(-\tilde{A}+\frac{l}{a}\tilde{B}^2-2\frac{l}{a}\tilde{A}^2
-\frac{l}{a}\tilde{B}-\tilde{A}\tilde{B})\right],\nonumber\\
L_2&=&\frac{1}{[(1-\tilde{B})^2-4\tilde{A}^2]}\left[\iota
J\tilde{A}- \frac{\iota
qg}{2}(1-\tilde{B}+2\frac{l}{a}\tilde{A})\right],
\end{eqnarray*}
\begin{eqnarray*}
L_3&=&\frac{1}{2(1-2\tilde{A}-\tilde{B})}\left[\iota qg(\frac{l}{a}+1)-\iota J\right],\\
L_4&=&\frac{1}{2(1-\tilde{B}+2\tilde{A})}\left[\iota J+\iota
qg(-\frac{l}{a}+1)\right]
\end{eqnarray*}
and $I_3$ can be obtained as
\begin{eqnarray}
I_3&=&N_1\ln\left[\frac{1-x(\tilde{A}+
\sqrt{\tilde{A}^2+\tilde{B}})}{1-x(\tilde{A}-\sqrt{\tilde{A}^2+\tilde{B}})}\right]
\nonumber\\&+&N_2\ln\left[1-2\tilde{A}x-\tilde{B}x^2\right]+N_3\ln[1+x],
\end{eqnarray}
where $x=\cos\theta$ and
\begin{eqnarray*}
N_1&=&\left[\frac{\tilde{A}-\tilde{B}}{2\sqrt{\tilde{A}^2+\tilde{B}}
(1-\tilde{B}+2\tilde{A})}\right],\\
N_2&=&\left[\frac{1}{2(1-\tilde{B}+2\tilde{A})}\right],\quad
N_3=-\left[\frac{1}{(1-\tilde{B}+2\tilde{A})}\right].
\end{eqnarray*}
Equations (\ref{7}), (\ref{WJ}) and (\ref{10}) can determine the
value for $W(r,\theta)$ and hence the outgoing massive particles
action can be obtained. For $m=0$, this expression diminishes to the
massless particles action. Similarly, we can determine the action
for the incoming particles either massive or massless.

\section{Outlook}

In this paper, we have used semiclassical WKB approximation to study
tunneling continuum of charged fermions from a pair of electrically
and magnetically charged accelerating and rotating BHs, together
with NUT parameter and cosmological constant. It is found that the
tunneling probabilities ((\ref{12}) and (\ref{12A})) of outgoing
charged fermions do not depend upon fermion's mass but only its
charge. For the family of BH solutions, the corresponding Hawking
temperatures ((\ref{New1}) and (\ref{NewWWW})) depend upon mass,
acceleration, rotation parameters and NUT parameter as well as
electric and magnetic charges of the pair of BHs involving
cosmological constant. Equations for the spin-down case are of the
identical form as in case of the spin-up particles with the
exception of negative sign. For both cases, either massive or
massless, the Hawking temperature indicates that the spin-up and
spin-down particles are transmitted at the similar tunneling rate
\cite{rbmann1}.

This work is the generalization of our previous work \cite{S1a} by
adding $\Lambda$ in the family of BH solutions. We see from graphs
that these solutions lead to expanding BH solutions. We would like
to mention here that the cosmological constant can be
positive/negative in general. However, for the sake of positive
temperature, the cosmological constant must be positive for this set
of parameters. We can take negative cosmological constant for some
other set of parameters but it is not sure whether it will give
temperature positive or negative. It is worth mentioning here that
for the PD family of BH solutions, in the absence of the
cosmological constant, all results reduce to the results already
given in \cite{S1a}.

The graphical representation (Figures \textbf{1} and \textbf{2})
indicates that whether the cosmological constant increases or
decreases, it has no effect on the horizon radius $r_{1+}$. However,
the horizon radius always increases (hence approaches to its maximum
value) whether $e$ is decreasing to zero or increasing to zero.
Similarly, the horizon radius always decreases (hence approaches to
its minimum value) whether positive $e$ is increasing to $+\infty$
or negative $e$ is decreasing to $-\infty$. Hawking temperature
remains positive for this choice of parameters. For $e<0$, the
temperature decreases as $e$ decreases (independent of $\Lambda$).
For $e>0$, the temperature increases as $e$ decreases. For $e=0$,
temperature approaches to its maximum value, while $\Lambda$ behaves
constantly. We see from Figures \textbf{3} and \textbf{4} that for
increasing $\Lambda$, the temperature $T_H$ and radius $r_{2+}$
increase when $e<0$, while these decrease when $e>0$. For $e=0$, the
Hawking temperature and horizon radius attain its maximum values
with positive increasing $\Lambda$.

The graphical behavior of horizons show that $r_{1+}$ is the outer
horizon, while $r_{2+}$ is the cosmological horizon. In the
tunneling picture, particles can also tunnel from the cosmological
horizon like the event horizon. The tunneling behavior is different
for these two horizons. The event horizon decreases when +ve-energy
particles tunnel across it, while the cosmological horizon expands.
The emitted particles are found to tunnel into the cosmological
horizon in the form of radiation \cite{A}-\cite{rep8}.

For de Sitter BHs, particles can be created at event horizon as well
as at cosmological horizon. At the event horizon, $+$ve-energy
(outgoing) particles tunnel out the BH horizon to form Hawking
radiation and $-$ve-energy (incoming) particles can fall into the
horizon along classically permitted trajectories. At the
cosmological horizon, outgoing particles can fall classically out of
the horizon and incoming particles tunnel inside the horizon to form
Hawking radiation for distant observer. Thus, the tunneling
probability of incoming particles through cosmological horizon
$r_{2+}$ of PD BHs can be written as
\begin{eqnarray}
\Gamma&=&\frac{\textmd{Prob}[\textmd{in}]}
{\textmd{Prob}[\textmd{out}]}= \exp[-4\textmd{Im}R_-],\nonumber\\
&=&\exp\left[\frac{4\pi\left[r_{2+}^2+(a+l)^2\right]\left(E-\Omega_HJ-
\frac{qer_{2+}}{\left[r_{2+}^2+(a+l)^2\right]}\right)}{\left(r_{2+}-r_{1+}\right)
\left(r_{2+}-r_{1-}\right)\left(r_{2+}-r_{2-}\right)}\right].\nonumber
\end{eqnarray}
By comparing the above expression with the Boltzmann factor, there
is only one choice to consider $-$ve-energy at cosmological horizon.
Thus, in de Sitter space massive particles have $-$ve-energy which
can tunnel inside the cosmological horizon. The Hawking temperature
at the cosmological horizon of PD BHs by using $-$ve-energy
particles can be written as given by Eq.(\ref{NewWWW}). This
temperature is same as for the temperature of outgoing particles.

Figures \textbf{1} and \textbf{3} show that the Hawking temperatures
at horizon radii $r_{1+}$ and $r_{2+}$ exhibit $+$ve behavior. Thus
the Hawking temperatures (\ref{New1}) and (\ref{NewWWW}) must be
$+$ve for outgoing particles through the event horizon and incoming
particles through the cosmological horizon. The graphical behavior
of temperature helps to know about the horizon radii of PD BH. These
verifications are consistent with already available in the
literature \cite{WJ3}.

Finally, it is pointed out that BHs with NUT charge are not
consistent with the existence of fermions insofar as such spacetimes
do not support spin structures \cite{n}. Here we have given
calculations that surprisingly show good agreement with known
results about the Hawking temperature in the limits in which they
apply. This agreement is of particular interest even when no spin
structure of $\Psi$ exists.

\vspace{0.25cm}

{\bf Acknowledgement}

\vspace{0.25cm}

We would like to thank the Higher Education Commission, Islamabad,
Pakistan, for its financial support through the {\it Indigenous
Ph.D. 5000 Fellowship Program Batch-IV}. We also appreciate the
referee's comments and providing a particular reference.

\end{document}